\documentclass[twocolumn,showpacs,prc,superscriptaddress]{revtex4}          
\newcommand{ \be }{\begin{equation}}       
\newcommand{ \ee }{\end{equation}}       
\newcommand{ \bea }{\begin{eqnarray}}       
\newcommand{ \eea }{\end{eqnarray}}       
\newcommand{ \la }{\langle}       
\newcommand{ \ra }{\rangle}       
\newcommand{ \eps }{\varepsilon}       
       
\newcommand{ \mean }[1]{\langle #1 \rangle}   
\newcommand{ \avg }[1]{\langle #1 \rangle}   
\newcommand{ \etal }{{\it et al.}}   
\usepackage{color}

\usepackage{graphicx} 
\usepackage{dcolumn} 
\usepackage{bm} 
\def \vobs {v_2^{\rm obs}} 
\def \sobsk {s_{2;k}^{\rm obs}} 
\def \cobsk {c_{2;k}^{\rm obs}} 
\def \ck {c_{2;k}} 
\def \sk {s_{2;k}} 
\def \ckp {c_{2;k}'}
\def \skp {s_{2;k}'}
\def \psirp {\Psi_{\rm RP}}
\def \psirpk {\Psi_{{\rm RP};k}}
\def \psia {\psi_a}
\def \psib {\psi_b} 
\def \psiak {\psi_{a;k}}
\def \psibk {\psi_{b;k}} 
\def \dpsiak {\Delta \psi_{a;k}}
\def \dpsibk {\Delta \psi_{b;k}} 
\def \dpsiabk {\Delta \psi_{ab;k}} 
\def \distrib#1 {\{#1\} }
\def \E#1{\mean{#1}}

\def \ES#1{\mean{#1}^2}
\def \sgmS#1{\sigma^2\{#1\} }

\begin{document}          
\title{       

A method for determining event-by-event elliptic flow fluctuations \\
based on the first-order event plane in heavy-ion collisions
} 

\author{Gang Wang}
\affiliation{University of California, Los Angeles, California 90095}
\author{Declan Keane}
\affiliation{Kent State University, Kent, Ohio 44242}
\author{Aihong Tang}
\affiliation{Brookhaven National Laboratory, Upton, New York 11973}
\author{Sergei A. Voloshin}
\affiliation{Wayne State University, Detroit, Michigan 48201}

\begin{abstract}     
A new method is presented for determining event-by-event fluctuations 
of elliptic flow, $v_2$, using first-order event planes.
By studying the event-by-event distributions of $v_2$ observables 
and first-order event-plane observables, average flow 
$\avg{v_2}$ and event-by-event flow fluctuations 
can be separately determined, making appropriate allowance 
for the effects of finite multiplicity and non-flow. 
The method has been tested with Monte Carlo simulations.
The connection between flow fluctuations and fluctuations of 
the initial-state participant eccentricity is discussed. 
\end{abstract} 
 
  
\maketitle

\section{Introduction}  

In heavy-ion collisions, the azimuthal distributions of emitted particles 
can be decomposed with a Fourier expansion~\cite{Methods}:
\be
\frac{dN}{d\varphi}  = 
\frac{1}{2\pi}\{ 1 + \sum_{n=1}^{\infty} 2v_n \cos n(\varphi -\psirp) \}, 
\label{equ:Fourier_expansion}
\ee
where $\varphi$ denotes the azimuthal angle of the particle
and $\psirp$ is the reaction plane azimuth (defined by the impact parameter 
vector). 
The Fourier coefficients,  
\be
v_n = \langle \cos n(\varphi -\psirp) \rangle \,,  
\label{equ:Fourier_coefficient}
\ee
are referred to as anisotropic flow of the $n^{\rm th}$ harmonic.  The 
second harmonic, elliptic flow, carries information on the early stage of 
heavy-ion collisions, and has been extensively studied.
Event-by-event flow fluctuations 
\cite{STARlongFlow130,STARlongFlow200,
MillerSnellings,Mrow-Shuryak,Phobos06,SorensenQM06} 
are of considerable interest as they must be sensitive to the 
physics of the very early stages of the collision, and  
because any fluctuation observable has potential relevance for phase 
transition phenomena. 
Understanding flow fluctuations would also greatly improve 
anisotropic flow measurements that  
are currently dominated by systematic uncertainties in which flow fluctuations 
play a crucial role.  
At fixed centrality, initial-state fluctuations 
in the spatial anisotropy of the participant zone will cause flow 
fluctuations.  In addition to this 
inevitable source of fluctuations, there might be additional fluctuation 
contributions that could offer unique insights into dynamical details 
of the collision process at very early times 
(1 fm$/c$ and earlier) \cite{Mrow-Shuryak}.  
The large observed elliptic flow at 
the Relativistic Heavy Ion Collider (RHIC) points to 
a very short thermal equilibration time in the framework of hydrodynamic 
models.  This puzzling feature calls for further investigation and 
could have alternative explanations or might be explained by exotic 
phenomena \cite{Mrow93-97,KovTuchin02,KNV03,Shuryak03}.  
Improved methods to experimentally determine flow fluctuations would 
be an important step towards addressing some or all of the open issues 
discussed above.

Fluctuations in the shape of the initial participant region, 
and in particular, in the orientation of the region's principal axes
relative to the direction of the impact parameter, lead to a
non-trivial picture of anisotropic flow.
In this picture, the apparent flow at mid-rapidity might be
different in direction and magnitude from the real flow as measured
with respect to the reaction plane.  The method proposed in this paper 
is sensitive to such a difference, and allows fluctuations in the 
orientation of the principal axes of the participant region to be measured.

RHIC data hold much promise for the purpose of understanding elliptic 
flow fluctuations, since $v_2$ is large, while the statistical noise 
arising from finite multiplicity, that tends to obscure the dynamical 
fluctuations of interest, is smaller than at lower energies.  
Most flow analyses at RHIC to date have relied on the second-order 
event plane, whereas in the present study, a case is presented for 
utilizing the first-order event plane to determine the mean elliptic 
flow, and to isolate the sought-after dynamical 
fluctuations about that mean.     
In RHIC experiments, first-order event planes can be obtained, for 
example, via the ZDC-SMD (Zero Degree Calorimeter
Shower Maximum Detector)~\cite{ZDC} or the 
Forward TPC~\cite{FTPC-NIM} of the STAR detector.  
In the scenario envisaged here, the fluctuating anisotropies are 
based on measurements near mid-rapidity, while the first-order event 
plane determination utilizes detectors that are far removed in rapidity. 
Consequently, non-flow effects (defined as azimuthal correlations 
that may contribute to $v_n$ measurements, but which are unrelated to 
the reaction plane orientation, or more generally, are unrelated to the 
initial geometry of the system) are believed to be 
negligible using this method \cite{STARv1at62}.  

\section{Technique}

With two independent first-order event planes $\psia$ and $\psib$, 
elliptic flow can be determined with the help of the relations
\bea
\vobs &=& \la \cos (2 \varphi - \psia - \psib) \ra
\nonumber \\
&=& \la \cos(2\varphi - 2\psirp) \ra \, \la \cos(2\psirp - \psia - \psib) \ra 
\nonumber \\
&=& v_2 \, \la \cos(\psia-\psib) \ra \,,
\label{equ:v2_observables}
\eea
where the last factor, 
$\la \cos(\psia-\psib) \ra = \mean{\cos(\psia-\psirp)} \mean{\cos(\psib-\psirp)}$  
is the product of the two first-order event plane resolutions~\cite{Methods}.
The above is based on the assumptions 
that the two event planes are independent, and that 
the distributions of $\psia$ and $\psib$ with respect to the true 
reaction plane are symmetric.

We introduce two {\em event-by-event} quantities
\bea
\ck &=& \la \cos 2(\varphi - \psirpk) \ra \\ 
\sk &=& \la \sin 2(\varphi - \psirpk) \ra \,, 
\eea
where index $k$ denotes the $k^{\rm th}$ event and the average is taken
over all particles in that event.  Using the equality
\begin{widetext}
\bea
\cos(2 \varphi - \psiak - \psibk) 
 &=&  \cos [ 2( \varphi-\psirpk) - (\psiak-\psirpk) - (\psibk-\psirpk)
]  \nonumber \\
 &=&  \cos 2( \varphi - \psirpk) \, \cos (\Delta\psiak +\Delta\psibk) 
    + \sin 2( \varphi - \psirpk) \, \sin (\Delta\psiak +\Delta\psibk) ,
\eea
where $\Delta\psiak = \psiak- \psirpk$ and $\Delta\psibk = \psibk - \psirpk$ 
(and a similar expression for $\sin(2 \varphi - \psiak - \psibk)$),
and averaging over particles in the $k^{\rm th}$ event, one finds
\bea
\cobsk &=& \ck \cos (\dpsiak +\dpsibk) + \sk \sin (\dpsiak +\dpsibk)
           \label{equ:v2_obs_k} \\
\sobsk &=& \sk \cos (\dpsiak +\dpsibk) - \ck \sin (\dpsiak +\dpsibk)
           \label{equ:s2_obs_k} \,.
\eea
\end{widetext}
We further assume that the 
distribution of the first-order event planes 
$\distrib{\dpsiak + \dpsibk}$ is independent of 
distributions $\distrib{\ck}$ and $\distrib{\sk}$ . 
This assumption is usually not valid for second-order event planes, 
and this illustrates one of the advantages of the present approach.  
Assuming that 
$\distrib{\dpsiak}$ and $\distrib{\dpsibk}$ 
are both symmetric around zero and independent, the distributions
$\distrib{\dpsiak + \dpsibk}$ and 
$\distrib{\dpsiak - \dpsibk \equiv \dpsiabk}$ are identical. 
We discuss the validity of all of the above assumptions 
in more detail at the end of Section~\ref{sEps}.

From Eqs.~(\ref{equ:v2_obs_k}) and (\ref{equ:s2_obs_k}) one can calculate the
mean (now averaged over all events) and the mean square of
$\cobsk$  and $\sobsk$ \,:
\be 
\E{\cobsk} = \E{\ck} \, \E{\cos \dpsiabk} = v_2 \, \E{\cos \dpsiabk} \,,
\label{equ:Ecobsk}
\ee
\be
 \E{\sobsk} =0 \,, 
 \label{equ:Esobsk}
\ee
\be
   \E{(\cobsk)^2} = 
   \E{\ck^2} \, \E{\cos^2 \dpsiabk} +  \E{\sk^2} \, \E{\sin^2 \dpsiabk} \,, 
   \label{equ:EcobskS}
\ee
\be
   \E{(\sobsk)^2} =  
   \E{\ck^2} \, \E{\sin^2 \dpsiabk} +  \E{\sk^2} \, \E{\cos^2 \dpsiabk} \,. 
   \label{equ:EsobskS}
\ee
Conventionally, $\E{\cos \dpsiabk}$ in Eq.~(\ref{equ:Ecobsk})
is regarded as a correction for the event plane resolution. 
From the above equations, one finds relations
\be
 \E{\ck^2} +  \E{\sk^2} =  \E{(\cobsk)^2}+\E{(\sobsk)^2} \,,
\ee
\be
 \E{\ck^2} -  \E{\sk^2} = \frac{\E{(\cobsk)^2}-\E{(\sobsk)^2}}
                               {\E{\cos 2 \dpsiabk}}. \\
\ee

The fluctuations, $\sgmS{\ck} = \E{\ck^2}- \ES{\ck} $ and 
$\sgmS{\sk} = \E{\sk^2}$, each have several contributions:  
dynamical flow fluctuations, non-flow, and a statistical part 
that is related to finite event multiplicity,  
\be
 \sgmS{\ck} = \sigma_{\rm dyn}^2\{\ck \} + \frac{\delta}{2} + 
              \frac{ 1 + v_4 - 2\ES{\ck} - \delta } {2M} \,, 
\ee
\be
\sgmS{\sk} = \sigma_{\rm dyn}^2\{\sk \} + \frac{\delta}{2} + 
             \frac{ 1 - v_4 - \delta } {2M} \,, 
\ee
where $M$ denotes multiplicity, and $\delta$ stands for the non-flow contribution. 
The $v_4$ term in the above equations arises from setting $n = 2$ 
in the equalities
\bea
\langle \cos^2 n(\varphi - \psirp) \rangle &=& 0.5(1 + v_{2n}) \,,
\label{equ:mean_square_v} \\
\langle \sin^2 n(\varphi - \psirp) \rangle &=& 0.5(1 - v_{2n}) \,.
\label{equ:mean_square_s}
\eea
Note that $v_4$ is usually negligible compared to 1.  Also, 
the terms inversely proportional to multiplicity can be
experimentally measured by studying the dependence of $\sgmS{\ck}$ and
$\sgmS{\sk}$ on the multiplicity of particles used in the event. 
Either measuring these terms, or just neglecting the difference in
$\sim 1/M$ terms
for $\sgmS{\ck}$ and $\sgmS{\sk}$ and evaluating the difference 
 $\sgmS{\ck} - \sgmS{\sk}$, one gets access to the difference 
$ \sigma_{\rm dyn}^2\{\ck \} -  \sigma_{\rm dyn}^2\{\sk\}$.
The latter is directly related to flow fluctuations, 
but as shown below, it also depends
on fluctuations in the orientation of 
the principal axes of the participant region with respect to the direction
of the impact parameter. 
We return to this question after discussing 
eccentricity fluctuations in Section III.

Relations established above have been tested in a Monte Carlo simulation.
In each event, the azimuthal angle of each particle at mid-rapidity 
has been assigned randomly according to the distribution of 
Eq.~(\ref{equ:Fourier_expansion}).
In this Monte Carlo simulation, only $v_2$ is non-zero, and 
non-flow effects have not been implemented.
From event to event, the $v_2$ value fluctuates according to a Gaussian distribution.
The first-order event plane follows a typical event plane distribution described in
Ref~\cite{Methods}.
In the first set of simulations, the input corresponds to
 5\% mean elliptic flow and 3\% dynamical flow fluctuations.
We set each of the first-order event plane resolutions to be 20\%, which
corresponds to $\avg{\cos\dpsiabk} = 4\%$.  Five multiplicities are tested:
$M=$ 25, 50, 100, 200 and 400.  In each case, ten million events were generated.
Table~\ref{tbl:test_multiplicity} lists the output results.
\begin{table}[hbt]
\begin{center}
\begin{tabular}{c|c|c} \hline \hline
Multiplicity  & $\E{\ck}  (\%)$        &  $\sigma_{dyn}\{\ck \} (\%)$
\\ \hline
        $25$  & $4.90 \pm 0.12$         &  $3.16 \pm 0.18$
\\ \hline
        $50$  & $4.97 \pm 0.09$         &  $3.00 \pm 0.14$
\\ \hline
       $100$  & $4.90 \pm 0.07$         &  $3.14 \pm 0.11$
\\ \hline
       $200$  & $4.96 \pm 0.06$         &  $3.03 \pm 0.10$
\\ \hline
       $400$  & $5.05 \pm 0.05$         &  $2.94 \pm 0.09$
\\    \hline \hline
\end{tabular}
\end{center}
\caption{
Results for the extracted mean elliptic flow for an input of 5\%, and its
dynamical fluctuation for an input of 3\%, based on simulated events
with different multiplicities.
The errors are statistical. }
\label{tbl:test_multiplicity}
\end{table}
Over a broad range of multiplicities, the output results well agree with the input
values within the statistical errors of the simulation.

\begin{figure}[t]
  \includegraphics[width=0.50\textwidth]{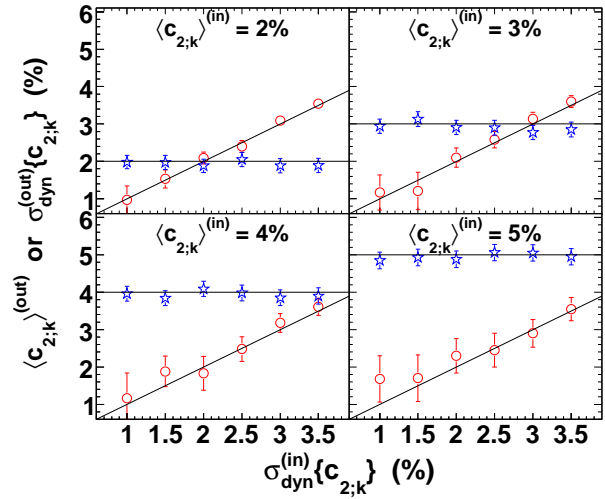}
  \caption{(color online)
The reconstructed mean $v_2$ (stars) and its dynamical fluctuation (circles)
as functions of input $\sigma_{\rm dyn}\{\ck \}$ for Monte Carlo events with
multiplicity of 100. The errors are statistical.
Panels are shown for input $\E{\ck}$ ranging from 2\% to 5\%.  
The horizontal lines represent the input mean $v_2$, and the diagonal lines
represent the input dynamical fluctuation of $v_2$.
}
\label{fig:simulation}
\end{figure}

We have also explored the robustness of the method with variations in the input
values of $\E{\ck}$ and $\sigma_{\rm dyn}\{\ck \}$.  In this second group of tests,
the multiplicity was fixed at $M = 100$, and one million events were generated in
each case.  The reconstructed mean $v_2$ and the
extracted dynamical fluctuations, $\sigma_{\rm dyn}\{\ck \}$,  
are shown in Fig.~\ref{fig:simulation},
and they are found to be consistent with the input values.

\section{Eccentricity fluctuations} \label{sEps}

\begin{figure}[t]
  \includegraphics[width=0.50\textwidth]{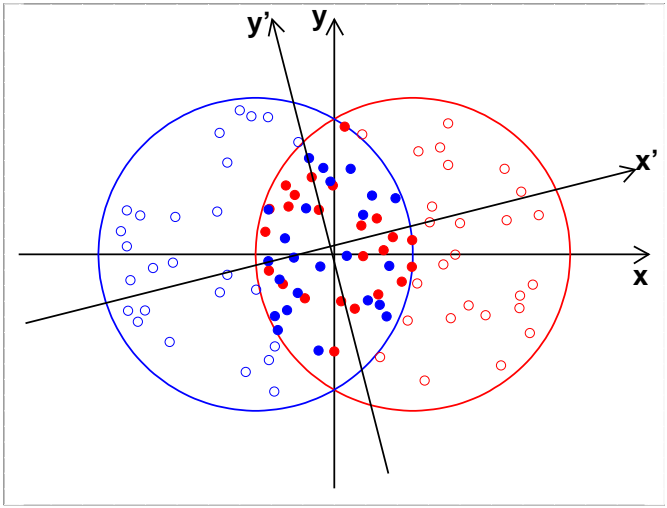}
  \caption{(color online) 
Schematic representation, in the plane transverse to the beam 
($z$) direction, of a collision between two identical nuclei. 
The $x$- and $y$-axes are drawn as per the standard convention. 
The solid circles illustrate a possible configuration of the 
participant nucleons.  Due to the fluctuation of this event, the 
overlap zone is shifted and tilted with respect to the $(x, y)$ 
frame.  $x'$ and $y'$ are the principal axes of inertia of the 
solid circles.
} 
\label{fig:tilt}
\end{figure}

In heavy-ion collisions, due to the finite number of participants, the
center of the overlap zone 
can be shifted and the orientation of the principal axes of the
interaction zone 
can be rotated with respect to the conventional coordinate system with
the $x$ axis pointing along the impact parameter, 
as illustrated in Fig.~\ref{fig:tilt}.  
As a result, the final particle distribution is symmetric 
about the $x'$ axis, instead of the $x$ axis.  
The effect of the shift is negligible compared to the effect of the 
rotation, and therefore we concentrate on the latter.  
The conclusion that the shift can be neglected is based on 
detailed Monte Carlo calculations using a Glauber model where the 
geometry of the interaction zone is defined by the position of the 
participating nucleons.

We define the angle between the $x'$ and $x$ directions 
to be $\Delta \psi_{\eps;k}$ for the $k^{\rm th}$ event. 

The standard definition of eccentricity~\cite{MillerSnellings} for 
the $k^{\rm th}$ event is 
\be
\eps_k = \frac{ \mean{y_i^2 - x_i^2}_k } 
              { \mean{y_i^2 + x_i^2}_k }       
\label{equ:eccentricity_definition}
\ee
where $x_i$ and $y_i$ denote the $i$-th participant coordinates and the
average is taken over all participants in the event.  
The coordinates $(x, y)$ and $(x', y')$ are linked by the rotation 
through $\Delta \psi_{\eps;k}$,
\bea
x_k  &=&  x'_k \cos \Delta \psi_{\eps;k} - y'_k \sin \Delta\psi_{\eps;k}
                    \nonumber \\
y_k  &=&  x'_k \sin \Delta \psi_{\eps;k} + y'_k \cos \Delta\psi_{\eps;k} \,,    
\label{equ:rotation}
\eea
which leads to the relation
\be
\eps_k = \eps'_k \, \cos 2 \Delta \psi_{\eps;k} \,.
\label{equ:eccentricity_relationship}
\ee
Note that from the definition of the rotated frame, $\mean{x_i' \, y_i'} = 0$.
The elliptic anisotropy in a given event is developed 
in the ${x',y'}$ plane 
such that
$\E{\skp}=0$ and $\sigma_{\rm dyn}\{ \skp \}=0$. 
Also, similar to the relation between eccentricities, one finds
\be
\E{\ck} = \E{\ckp} \, \E{\cos 2 \Delta \psi_{\eps;k}} \,,
\ee
which follows directly from 
\bea
\ck   &=& \ckp  \cos 2 \Delta \psi_{\eps;k} - \skp \sin 2 
\Delta \psi_{\eps;k} \,,       
\\
\sk &=& \ckp \sin 2 \Delta \psi_{\eps;k} + \skp \cos 2 
\Delta \psi_{\eps;k} \,.          
\label{equ:s2_tilt}
\eea
Therefore $\E{\ck}$ is always less than or equal to $\E{\ckp}$.  

Note that 
\bea
\ck^2 + \sk^2 = \ckp^{\,2} + \skp^{\,2} \,,   
\label{equ:tilt_square_sum}
\eea
which reflects the fact that such a combination
is independent of the plane in which it is calculated, as it
depends only on the particle pair angle differences in the event.
As mentioned in the previous section, this sum contains the 
flow fluctuation contribution (including the fluctuation in the
orientation of the principal axes of the participant zone) 
as well as the non-flow contribution. 
The non-statistical part of this sum (everything except $\sim 1/M$ terms)
corresponds exactly to $v_2^2\{2\}$ --- elliptic flow measured with
two-particle correlations at mid-rapidity.
To remove non-flow contributions, one should consider the difference
\be
  \E{ \ck^2 - \sk^2} 
= \E{ \ckp^{\,2} - \skp^{\,2} } \, \E{ \cos 4 \Delta \psi_{\eps;k} } .
\label{equ:tilt_square_dif_prime}
\ee
The non-statistical part of this difference,
\be
\E{ \ck^2 - \sk^2}_{\rm dyn} =
\E{\ckp^{\,2}}_{\rm dyn} \, \E{ \cos 4 \Delta \psi_{\eps;k} } ,
\ee
provides an important relation between fluctuations measured with
respect to the first-order reaction plane, $\E{\ck^2 - \sk^2}_{\rm dyn}$,
flow fluctuations measured at mid-rapidity (which includes
effects of the fluctuations in the geometry of the participant zone), 
and the distribution in $\Delta \psi_{\eps;k}$.

In the picture described above, when the flow fluctuations are driven
by fluctuations in the participant eccentricity, 
it is not obvious that 
the two first-order event planes
defined by spectators from the two nuclei are independent, 
nor is it obvious that 
the first-order event plane is independent of the second-order
event plane defined by the participants. Indeed, the positions of the
spectators are somewhat correlated with the positions of the participants, 
but as we found using the Monte Carlo Glauber model, this has a negligible 
effect on correlations of the event planes.  In our study, we used the center of
gravity of the spectator distribution with respect to 
the nuclear center
to define the first-order event plane for each nucleus ($\psi_{1,a}$ and $\psi_{1,b}$),
and the second-order event plane ($\psi_{2}$) was defined by the minor axis of the 
participant zone.  Using the center of the collision instead of the center of
gravity of the spectator distribution would make the correlation effects 
even smaller.

We find that for most centralities, the correlations 
\be
{ {\la \cos(\psi_{1,a}-\psi_{1,b}) \ra} \over
\mean{\cos(\psi_{1,a}-\psirp)} \,
\mean{\cos(\psi_{1,b}-\psirp)} }
 - 1 
\ee 
and 
\bea
 {\la \cos(\psi_{1,a} + \psi_{1,b} - 2 \psi_2) \ra} \over
\mean{\cos(\psi_{1,a}-\psirp)} \,
\mean{\cos(\psi_{1,b}-\psirp)} \,
\mean{\cos(2\psi_2 - 2\psirp)}  \nonumber \\
 - 1 ~~~~~~~~~~~~
\eea 
are at the sub-percent
level, with a maximum of about a few percent for the 5\% most central collisions.
Besides the event plane correlations due to
the correlated positions of spectators and participating nucleons,
momentum conservation also deserves consideration. 
We refer here to an experimental study~\cite{STARv1at62} 
which concluded that there is negligible momentum-conservation correlation between 
the event plane based on spectators from each nucleus separately and the orientation 
of directed flow close to mid-rapidity.

\section{Discussion and summary}

It is useful at this stage to consider the notation
$\E{\ckp} \equiv \E{v_2'}$ and $\E{\ckp^{\,2}}_{\rm dyn} = \E{v_2'^{\,2}}$ \,, 
where $v_2'$ is the ``apparent'' flow at mid-rapidity --- elliptic event
anisotropy measured with respect to the principal axes of the participant
zone.  
Much progress towards a direct measurement of flow fluctuations at 
mid-rapidity has been reported recently~\cite{Phobos06,SorensenQM06}. 
Then the above equations can provide important information on fluctuations
in the orientation of the principal axes of the participant zone.

Note that the orientation of the participant zone (and,
consequently, the apparent anisotropic flow) can depend on the rapidity of
the particles under study. Then, in principle, one can study the
correlations in the orientation of anisotropic flow as function of
particle rapidity. Such information will be very valuable for the
reconstruction of the initial conditions in heavy-ion collisions.

In summary, various suggestions in the literature point to dynamical 
event-by-event fluctuations in elliptic flow as being of great interest 
in the realm of RHIC physics, and such fluctuations are argued to be
 especially relevant 
for understanding collision dynamics at the earliest times.  On a more 
technical level in experimental methodology, the magnitude of flow 
fluctuations associated with current elliptic flow measurements is not 
understood, and this uncertainty affects the overall systematic error on 
these measurements.  Prompted by the above considerations, this work 
presents a new method for experimental analysis of elliptic flow in a 
scenario where the first-order event plane can be resolved.  
The method 
allows the extraction of mean $v_2$ and its dynamical event-by-event 
fluctuations, and good immunity to both statistical fluctuations and 
non-flow effects can be expected. 
Simulations have been presented that
validate the method under a range of conditions similar to those observed
in RHIC data.  
It has been shown that measurements of flow fluctuations using the
first-order event plane, accompanied by measurements of apparent flow
fluctuations at mid-rapidity, can also provide important information on
the fluctuations of the participant zone.

\begin{acknowledgments}
       
We acknowledge useful discussions with Art Poskanzer, Raimond Snellings and 
Paul Sorensen, and we thank them for their helpful suggestions.   
\end{acknowledgments}

  
\end{document}